\documentclass[trackchanges,twocolumn]{aastex701}
\usepackage{footnote}

\begin{document}

\title{Formation of Cosmic Noon Protogalaxies via Quasar-Induced Fragmentation of a Cosmic Filament}

\author[orcid=0000-0003-3726-5611]{Marko Mi\'ci\'c}
\affiliation{Homer L. Dodge Department of Physics and Astronomy, The University of Oklahoma, Norman, OK, USA}
\email[show]{micic@ou.edu}  

\author[]{Themiya Nanayakkara} 
\affiliation{Centre for Astrophysics and Supercomputing, Swinburne Institute of Technology, Hawthorne, VIC, Australia}
\email{fakeemail2@google.com}

\author[gname=Savannah,sname=Africa]{Xinyu Dai}
\affiliation{Homer L. Dodge Department of Physics and Astronomy, The University of Oklahoma, Norman, OK, USA}
\email{fakeemail3@google.com}
\author{Jeremy Bailin}
\affiliation{Department of Physics and Astronomy, University of Alabama, Tuscaloosa, AL, USA}
\email{fakeemail4@google.com}
\author{Miljan Kol\v ci\'c}
\affiliation{PajdoX, Belgrade, Serbia}
\email{fakeemail4@google.com}

\begin{abstract}

When black hole jets encounter ambient medium, they can compress the gas, trigger star formation, and create stellar clusters containing tens of thousands of stars. Here, we report a remarkable discovery of such a phenomenon that happened just 2.2 billion years after the Big Bang, during the Cosmic Noon era. Quasar SDSSJ141924.44+532315.5, powered by a one-billion-solar-mass black hole, is seen blasting a powerful jet that interacts with a hypermassive gas reservoir, creating a fascinating, clumpy, arc-like structure spanning over 250 kiloparsecs in projected length, consisting of at least eight clumps. Each clump contains billions of stars, is as massive as the Milky Way, and exhibits extreme levels of star formation. We interpret these findings as fragmentation of a cosmic filament triggered by a jet overpressurized expanding cocoon, which leads to the birth of protogalaxies, a process observed at scales never seen before. We find that the physical conditions within the filament are favorable for a fragmentation scenario to occur. We also discuss the survivability and evolution of individual clumps in the context of unsolved galaxy formation theory problems.
\end{abstract}



\section{Introduction} 

Compact objects, such as black holes or neutron stars, are known to be able to launch collimated outflows of ionized matter. The launching mechanisms of these outflows or jets likely originate from dynamical interactions within the fast-spinning accretion disk, which create chaotic magnetic fields. The charged particles from the accretion disk are then ejected, sometimes close to the speed of light, along the axis of rotation \citep{2019ARA&A..57..467B}. The energetic jets are commonly observed at X-ray and radio wavelengths and sometimes extend to monumental sizes.

As jets travel through space, they often interact with the ambient medium. The jet-medium interaction affects the morphology of the jet by leading to the formation of compact, bright knots and hotspots, extended radio lobes, or causing jet deflection and deviation from a linear trajectory. Similarly, the moving jet impacts the medium through which it is traveling. Relatively low-$z$ studies found that supermassive black hole jets can lead to high ionization of the surrounding gas, the formation of bow shocks, and outflows propagating perpendicular to the jet direction \citep{2024ApJ...977..156D}. Other studies have found that jets can sometimes trigger star formation in their surroundings, through positive feedback \citep{2015A&A...574A..34S,2022A&A...657A.114C}. Moreover, the origin of beads-on-a-string-like structures, consisting of many young stellar clusters, has been linked to the interaction of jets with massive gas clouds, which pushes the gas in lateral directions and triggers intensive, larger-scale episodes of star formation \citep{2022Natur.601..329S, 2024ApJ...963....1O}. Higher-$z$ studies of jet feedback, although scarce and limited, demonstrate that jets play a significant role in shaping and regulating the evolution of the ambient medium in early epochs as well \citep{2000ApJ...540..678B,2004ApJ...612L..97K,2023MNRAS.522.4548D}. 

This paper reports the discovery of a jet and a hotspot associated with a $z$=2.948 quasar SDSSJ141924.44+532315.5 (SDSSJ14, hereafter). Deep archival optical imaging revealed a 250-kpc-long beads-on-a-string-like structure perpendicular to the trajectory of the jet, consisting of at least eight clumps, with each clump being as massive as a stellar component of the Milky Way. We interpret these findings as a potential gargantuan quasar-induced fragmentation of a cosmic filament, resulting in the creation of protogalaxies.

The paper is organized as follows: Section 2 presents the methodology and data used in this work; Section 3 presents the main results; Section 4 interprets the results and discusses the impact of this discovery on various anomalies in galaxy formation theory.


\section{Methodology and Data} \label{sec:style}
\subsection{Chandra LOw Count Quasar Survey}
This discovery is part of a broader Chandra LOw-Count Quasar (CLOCQ) survey. We identified $\sim$20000 quasars observed with Chandra, primarily serendipitously, with most of them not being analyzed in X-rays. Even
though the main characteristics of quasars are high luminosities, due to their extreme distances,
shallow observational coverage, or different environmental effects, quasars and their energetic features
can sometimes appear faint in X-rays and go unnoticed. A careful, large-scale, energy-dependent,
low-count X-ray study of quasars has the potential to reveal a plethora of intriguing but
overlooked sources, including previously unknown jets, hotspots, heavily obscured quasars, dual quasars, and extended X-ray haloes.
\subsection{X-ray, radio, and optical data}
The Chandra X-ray telescope observed SDSSJ14 on two occasions in 2003 and 2006 (PI: Ellingson, PI: Garmire) using the ACIS-S instrument with a total exposure time of 57.45 kiloseconds. We used the Chandra Interactive Analysis of Observations (CIAO 4.17) tool and the calibration database CALDB 4.12.0 to perform the X-ray analysis \citep{2006SPIE.6270E..1VF}. The observations were downloaded, reprocessed, stacked, and filtered into three energy bands: soft (0.3-1 keV), medium (1-2 keV), and hard (2-6 keV). For the investigation of the radio properties of SDSSJ14, including its jet and lobes, we used data from the Very Large Array Sky Survey (VLASS) \citep{2020PASP..132c5001L}. For imaging of its environment, we used the Very Large Array FIRST image cutout server\footnote[1]{https://third.ucllnl.org/cgi-bin/firstcutout}. For optical imaging of the region, we used the Subaru Hyper-Suprime Cam Legacy Archive (HSCLA) archival data \citep{2021PASJ...73..735T}. Photometric information on SDSSJ14 and the sources in its immediate neighborhood, used for analysis throughout this paper, was obtained from the HSCLA public archive \footnote[2]{https://hscla.mtk.nao.ac.jp/doc/acknowledgement-hscla2016/}.
\begin{figure*}[h]%
\centering
\includegraphics[width=0.9\textwidth]{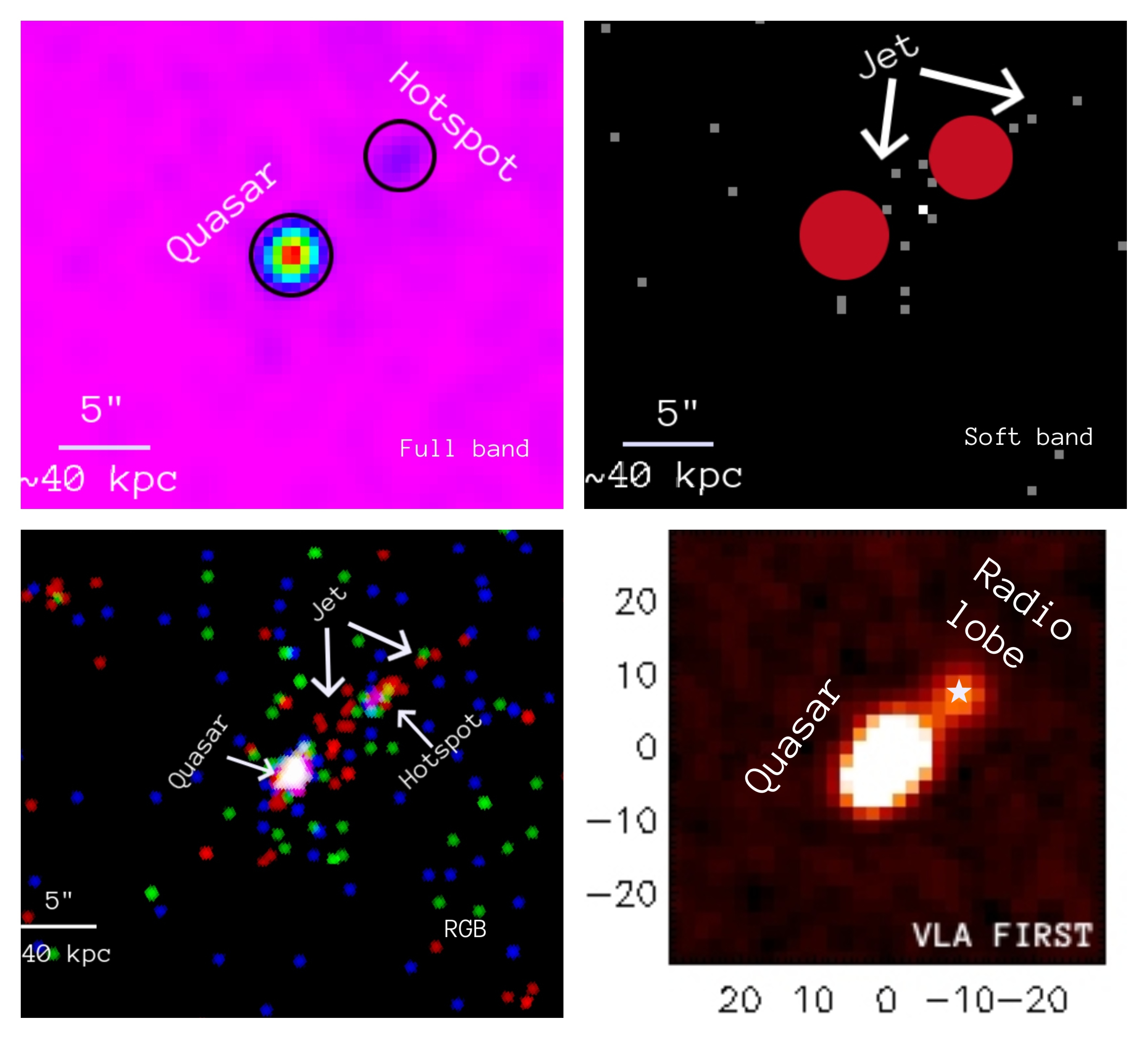}
\caption{X-ray and radio images of SDSSJ14. Top left: Full band Chandra image showing the quasar and hotspot. Top right: Soft band Chandra image shows the jet-like structure starting from the quasar and going through the hotspot. The quasar and hotspot are masked with red circles to enhance the jet’s visibility. Bottom left: Chandra HiPS RGB image (red-soft energy photons, green-medium energy photons, blue-hard energy photons). This image shows clear evidence of a jet-like structure traced by red dots. Bottom right: VLA FIRST radio image showing the central quasar and a radio lobe coinciding with the Chandra-detected hotspot, marked with a white star symbol. The x and y axes are showing RA and declination offset in arcseconds. All panels' color coding and smoothing are chosen to maximize the visibility of the feature of interest.}\label{fig1}
\end{figure*}
\begin{figure*}[h]%
\centering
\includegraphics[width=0.9\textwidth]{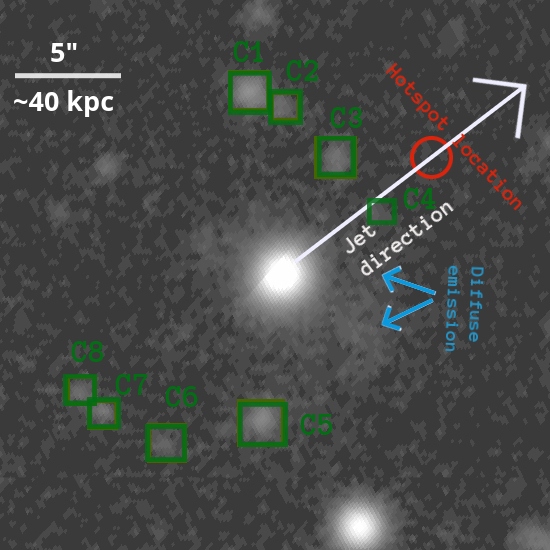}
\caption{HSCLA optical g-band image of SDSSJ14’s environment. The image is centered on the quasar. The rectangles label the locations of eight clumps within the arc (C1-C8). EAZY and FAST++ analysis was successfully performed for C1, C3, C5, and C7. The white arrow shows the direction of the X-ray-detected jet. The red circle shows the location of the X-ray- and radio-detected hotspot. The blue arrows point towards the diffuse, unresolved emission region between the two ends of the arc. North is up, and east is to the left.}\label{fig2}
\end{figure*}
\section{Results} \label{sec:results}
\subsection{X-ray and radio analysis}
SDSSJ14 was previously known as a strong X-ray emitter. At its location, we detected a powerful X-ray source with 272 net photons, corresponding to the luminosity of L$_X$=5$\times$10$^{45}$ erg s$^{-1}$. Moreover, Chandra observations revealed another, fainter source, detected with 25 net photons, located 8.3 arcseconds (or 65.5 kiloparsecs, assuming the same redshift) away from SDSSJ14. Upon closer inspection of the Chandra data, we discovered a 125 kiloparsec-long jet-like structure connecting the quasar and the mysterious secondary source, and extending beyond. The jet is significantly detected, but with only 10 net X-ray photons, all in soft energies under 1 keV, and it is only possible to observe it after energy filtering the Chandra data. 

This jet travels through space, interacts with the ambient medium, and creates a bubble of X-ray emission. This so-called hotspot manifests itself in Chandra imaging as the secondary X-ray source. Due to its inconspicuous nature, the jet remained unnoticed for 20 years after the observations were taken, and its low-count nature demonstrates the need for a dedicated large-scale survey to detect similar faint sources, such as the CLOCQ survey.

Additional evidence for the association of the quasar, the secondary mysterious X-ray source, and the extended jet-like X-ray source can be obtained from archival radio data. The VLASS survey detected two radio lobes associated with SDSSJ14, one of which spatially coincides with the location of the Chandra-detected secondary source (i.e., the potential hotspot). The radio lobes indicate interaction between the jet and the ambient medium, and often have compact hotspots embedded within them, implying they are being actively supplied with energy \citep{1984MNRAS.208..813S}. X-ray and radio imaging of the system is shown in Figure \ref{fig1}.
\subsection{EAZY analysis}
Archival HSCLA optical images of the region, the deepest available of this part of the sky, revealed a nearly continuous arc of emission. The arc consists of a central diffuse emission region that extends into two mirror-symmetrical arms, each containing four clumps, strikingly resembling local examples of beads-on-a-string structures. The HSCLA imaging is shown in Figure \ref{fig2}. We used grizY magnitude measurements from the HSCLA and EAZY software \citep{2008ApJ...686.1503B} to determine the photometric redshifts of the clumps. To perform the analysis, we used the \textsc{cww+kin} model, which is appropriate for star-forming galaxies \citep{1980ApJS...43..393C, 1996ApJ...467...38K}, and fitted all templates simultaneously. The redshift was successfully estimated in the four brightest clumps (C1, C3, C5, C7) with parameter \textsc{odds}$>$0.99. The \textsc{odds} parameter represents the redshift quality parameter and is defined as the fraction of the total integrated probability within $\pm$z of the photometric redshift estimate \citep{2000ApJ...536..571B}. All four clumps have redshifts comparable to the spectroscopic redshift of the quasar, suggesting a potential spatial relationship between the quasar and the arc. However, due to uncertainties in the photometric redshift measurements, a possibility of chance alignment remains. The results are summarized in Table 1. Adopting the accurately measured quasar’s spectroscopic redshift implies that the arc extends over 250 kiloparsecs in projected length.
\subsection{FAST++ analysis}
We further investigated the physical properties of the four brightest clumps using the FAST++ code\footnote[3]{https://github.com/cschreib/fastpp}, an improved modification of the FAST code \citep{2009ApJ...700..221K}. We assumed the Bruzual \& Charlot population synthesis model \citep{2003MNRAS.344.1000B}, Chabrier initial mass function \citep{2003PASP..115..763C}, delayed exponentially declining star formation histories, and Calzetti dust attenuation law \citep{1994ApJ...429..582C}, and the redshift was fixed at the redshift of SDSSJ14, $z$=2.948. The stellar masses of the four brightest clumps range from 1.5$\times$10$^{10}$ \(M_\odot\) to 6.2$\times$10$^{10}$ \(M_\odot\). Each clump potentially contains billions of stars and is as massive as the stellar component of the Milky Way or Andromeda. Additionally, we calculated that star formation rates range from 18 to 62 \(M_\odot\) yr$^{-1}$. These values are comparable to the star formation rates of main-sequence galaxies with similar stellar masses at similar redshifts \citep{2011ApJ...739L..40R}. However, it is worth noting that the FAST++ analysis was based solely on rest-UV data and can be significantly improved with the addition of rest-optical observations. Detailed information on the physical properties of the clumps is provided in Table 1. 
\section{Discussion}
\subsection{Interpretation}
Various morphological clues seen in Figure \ref{fig2} suggest a non-random arrangement and provide hints into the formation mechanism and dynamical evolution of the system. \begin{itemize}
    \item The nearly uniform clumpy arc propagates perpendicularly to the direction of the jet
    \item The arc consists of central, unresolved diffuse emission, extending into two clumpy arms.
    \item The two arms appear symmetrical. Each arm contains four clumps, the interior clumps exhibit almost constant spacing, and there is a build-up of terminal pairs of clumps on either side of the arc (C1-C2 and C7-C8)
    \item The jet does not pass through the center of the arc, and the hotspot lies downstream
\end{itemize}
The most likely physical mechanism responsible for the observed phenomenon is beads-on-a-string fragmentation of a cosmic filament compressed by the AGN cocoon. As the jet travels through space, it encounters the ambient medium, driving the formation of bow shocks and inflating a hot, overpressurized cocoon of shocked gas around its trajectory \citep{1989ApJ...345L..21B,2017MNRAS.472.4707B,2024ApJ...973..148D}. The cocoon expands laterally and exerts pressure on the preexisting filament of gas. This external pressure leads to the radial compression of the filament, funneling the material towards the center, increasing the central density, and decreasing the scale height. Radial collapse must be halted by an increase in temperature at higher densities or some other mechanism to allow fragmentation to occur \citep{2014prpl.conf...27A}. Once this condition is met, if the line mass of the filament exceeds the critical line mass, the filament can no longer be in hydrostatic equilibrium. The supercritical filament becomes favorable for the development and growth of longitudinal perturbations \citep{1992ApJ...388..392I,1997ApJ...480..681I}. As a result, the filament fragments into dense clumps, forming a beads-on-a-string-like structure. Linear analyses of self-gravitating isothermal cylinders predict fragmentation into more or less regularly spaced condensations driven by the fastest-growing mode \citep{1997ApJ...480..681I,2016MNRAS.458..319C}, a result supported by Herschel observations of quasi-regular core spacing in nearby filaments \citep[][and references therein]{2014prpl.conf...27A}. The clustering of terminal clumps on both ends of the arc is the telltale signature of beads-on-a-string fragmentation of finite filaments. The ends of filaments, having fewer confining forces, can experience a faster growth of instabilities. The so-called edge-driven collapse, a process predicted by theory but rarely observed, occurs when the fastest-growing unstable modes concentrate towards the edge and lead to the earlier formation of denser and closer clumps, as observed on either end of the SDSSJ14's arc \citep{2011ApJ...740...88P,2015MNRAS.449.1819C,2023MNRAS.521.5152H}. A consequence of the edge-driven collapse might be the existence of a central, diffuse emission region consisting of the unfragmented gas streaming inward.

The off-center jet-arc intersection and the downstream hotspot suggest that the jet is not directly responsible for filament fragmentation. If the jet caused direct, mechanical, or radiative triggering, the clumps would form along its path, tracing the jet propagation, and the maximum star formation would occur at the intersection point. Additionally, the hotspot marks the location where the jet mostly dissipates and is located downstream of the arc. Had the jet been directly responsible for the observed fragmentation, the maximum impact would be at the point of intersection. Instead, the jet passes through the filament, inflating the overpressured cocoon in lateral directions, which compresses the filament along its length, triggering fragmentation. 
\subsection{Physical properties of the filament}
The filament fragmentation into protogalaxy-sized clumps requires specific physical conditions, high densities, masses, and pressures. We calculated spacings between clumps, excluding the C4-C5 pair separated by the central diffuse region, and found an average fragmentation length of $\lambda_{frag}$=17.4 kpc. The fragmentation length $\lambda_{frag}$ is related to the isothermal scale height H as:
\begin{equation}
    \lambda_{frag}=22H
\end{equation}
implying a scale height of H=0.8 kpc. The scale height H is also related to the density of the medium via:
\begin{equation}
    H=\frac{c_s}{\sqrt{4\pi G\rho}}
\end{equation}
where c$_s$ is the speed of sound, G is the gravitational constant, and $\rho$ is density, which can be rewritten as $\rho$=n$\mu$m$_p$, where $n$ is the number density, $\mu$ is the mean molecular weight, and can be set at 0.6 for the ionized gas, and $m_p$ is proton mass \citep{1953ApJ...118..116C,1964ApJ...140.1056O}. Rearranging for $n$ and adopting a plausible range of speeds of sound of c$_s$=15-50 km s$^{-1}$ (depending on whether velocity dispersion due to turbulence and Alfven speed due to magnetic pressure are accounted for along with isothermal speed of sound) for a 10$^4$ K gas yields a number density of $n$=0.5-5 cm$^{-3}$. This value is several orders of magnitude higher than what is usually seen in high-$z$ filaments, and is consistent with the filament being compressed by the expanding AGN cocoon \citep{2010MNRAS.407..613G}.

We can estimate the lower limit of the line mass of the filament by dividing the apparent length of the filament by the derived stellar mass of four of the brightest clumps: $\frac{M}{L}>$ 5$\times$10$^8$ \(M_\odot\) kpc$^{-1}$. On the other hand, the critical line mass of an isothermal cylinder in equilibrium is given as:
\begin{equation}
    (\frac{M}{L})_{crit}=\frac{2c_s^2}{G}
\end{equation}
which yields values of 1$\times$10$^8$ to 1$\times$10$^9$ \(M_\odot\) kpc$^{-1}$. The observed line mass likely exceeds the critical line mass, implying that the conditions within the filament are susceptible to longitudinal instabilities that lead to the fragmentation of the filament. This process has been observed in the case of galactic filaments, although at much smaller scales \citep[e.g.,][]{2010A&A...518L.102A}.
\subsection{Survivability and evolution of clumps}
Do the clumps remain bound to the quasar, or do they break free? Do they evolve into regular galaxies, or do they disperse? This is a non-trivial question to answer because their survivability and evolution depend on the environment, the dark matter content of the gas reservoir from which they are born, their internal structure, and the physical properties of the AGN and jet.

The major unknown quantity in this analysis is the dark matter content of the clumps. We calculate potential evolutionary outcomes considering two cases: 1) dark-matter-free clumps, larger analogs of tidal dwarf galaxies, with mass $m_{c}$=2$\times$10$^{10}$ \(M_\odot\); 2) each clump resides in its own dark matter subhalo (i.e., bonafide protogalaxies) with mass $m_{cDM}$=2$\times$10$^{11}$ \(M_\odot\). In both cases we assume approximate effective radius of $r_{eff}$=3 kpc, distance from the quasar of $R$=50 kpc, and the host halo mass of $M_{Host}$=10$^{12}$ \(M_\odot\). The SDSSJ14 environment is relatively isolated, and the closest neighbor confirmed spectroscopically is located at a distance $\sim$30 Mpc, implying that no external forces are exerted on the clumps. 
\subsubsection{Tidal stripping} The probability of the clumps being disrupted by the tidal field of the host can be estimated by calculating their tidal radius:
\begin{equation}
    r_t=R(\frac{m_{c/cDM}}{3M_{Host}})^{1/3}
\end{equation}
resulting in r$_t \approx$10 and 20 kpc, for the dark-matter-free and dark matter subhalo scenarios, respectively. The tidal radii of the clumps are comfortably larger than the assumed effective radius, making them stable against the host tidal stripping regardless of their dark matter content. However, this is only a simplified picture since clump-clump interactions or mass loss through star-formation-driven winds can unbind them.

\subsubsection{Dynamical friction} The dynamical friction infall time can be estimated using the following expression:
\begin{equation}
    t_{fric}=\frac{1.17}{ln \Lambda}\frac{R^2v_c}{Gm_{c/cDM}}
\end{equation}
where the Coulomb logarithm can be approximated as ln$\Lambda \approx$ 3 \citep{2008MNRAS.383...93B}, and $v_c$ is the circular velocity of a clump at a given radius R, and based on the initial conditions can be set at $v_c$=200 km s$^{-1}$. The resulting dynamical friction timescales are $t_{fric} \approx$ 2.5 and 0.25 Gyr, for the dark-matter-free and dark matter subhalo scenarios, respectively. Dark-matter-free clumps are longer-lived structures that can persist to create a bound satellite population, the future bright central galaxy of a cluster, or a fossil group, given the apparent isolation of SDSSJ14. On the other hand, the timescale for clumps residing in dark matter subhaloes is significantly shorter. They are likely to spiral in and merge with the central object, allowing the central galaxy to accrete a monumental amount of mass. Interestingly, there are a few examples of overmassive galaxies that lie below the baryonic Tully-Fisher relation, such as NGC 1961 or UGC 12591. X-ray observations of these galaxies suggest that most of their stellar mass is not assembled through the cooling of the gas halo, but rather via some other unknown mode of accretion \citep{2011ApJ...737...22A,2012ApJ...755..107D}. This discovery provides a plausible explanation for the origin of such galaxies.

\subsubsection{Ram-pressure stripping} \cite{1972ApJ...176....1G} formulated that the gas remains retained within the galaxy as long as the ram-pressure stripping does not exceed the gravitational potential restoring forces:
\begin{equation}\label{eq6}
 P_{ram}=\rho_{CGM}v^2\geq2\pi G\Sigma_{tot}\Sigma_{g}
\end{equation}
where P$_{ram}$ is the ram pressure, $v$ is the speed of the clump traveling through the circumgalactic medium of density $\rho_{CGM}=\mu m_pn_{CGM}$, $\Sigma_g$ is the gas surface density, and $\Sigma_{tot}$ is the total restoring surface density. For the baryon-only case, assuming the gas fraction of 0.5, the total restoring surface density can be rewritten as $\Sigma_{tot}=\Sigma_{*}+\Sigma_{g}=2\Sigma_{g}$. Substituting in the Equation \ref{eq6} and solving for $n_{CGM}$ results in a minimum required density of the circumgalactic medium of $n_{CGM}$=4 cm$^{-3}$. The dark matter subhalo scenario requires even higher densities since the total restoring surface density increases due to the dark matter contribution. These values exceed typical CGM densities by several orders of magnitude, indicating that the clumps are practically immune to ram pressure stripping in both scenarios \citep{2021MNRAS.500.1476H}. 

\subsubsection{Can clumps escape?}

The escape velocity from the host halo at a given radius can be expressed as:
\begin{equation}
    v_{esc}=\sqrt{\frac{2GM_{Host}}{R}}
\end{equation}
resulting in $v_{esc} \approx$ 415 km s$^{-1}$. The kinetic energy necessary to accelerate the clump of mass $m_c$ or $m_{cDM}$ to $v_{esc}$ is:
\begin{equation}
    E=\frac{m_{c/cDM}v_{esc}^2}{2}
\end{equation}
which results in $E\approx$ 4.5$\times$10$^{58}$ and 4.5$\times$10$^{59}$ erg, for the dark matter free and the dark matter subhalo clump, respectively. The jet kinetic power needed to deliver the energies derived above, during a typical AGN episode of 10$^5$ years \citep{2015MNRAS.451.2517S} is $P_{jet}$=1.4$\times$10$^{46}$-1.4$\times$10$^{47}$ erg s$^{-1}$. 

The jet power can be calculated using the relation from \citep{1999MNRAS.309.1017W}:
\begin{equation}
    P\approx f^{3/2}3\times10^{38}(\frac{L_{151}}{10^{28} W Hz^{-1} sr^{-1}})^{6/7}W
\end{equation}
where $f$ is the parameter that accounts for systematic errors in model assumptions and usually has a value in the range of $f$=1-20, and $L_{151}$ is the 151 MHz luminosity of the radio lobe. After converting the observed 3-6 GHz integrated flux of radio lobe \citep{2021ApJ...914...42B} to 151 MHz luminosity by assuming the spectral index of $\alpha$=0.8, and adopting $f$=10, we found jet power of $P_{jet}\approx$ 2$\times$10$^{46}$ erg s$^{-1}$. However, with a different choice of $\alpha$ and $f$ this value can fluctuate by an order of magnitude. Additionally, only a small fraction of the jet power will go into the bulk acceleration of clumps, and the efficiency of clump acceleration significantly decreases with increasing clump mass \citep{2012ApJ...757..136W}. Therefore, clumps residing in dark matter subhalos will remain bound to the quasar, but there is a reasonable possibility that dark matter-free clumps may escape. If they manage to escape and survive, they might manifest themselves in the present day as rare and mysterious large, dark matter-deficient galaxies \citep{2018Natur.555..629V,2019ApJ...874L...5V}.
\subsection{Alternative scenarios} 
We consider two alternative explanations:
\begin{itemize}
    \item The arc is a gravitationally lensed background object
    \item The arc and quasar are chance-aligned
\end{itemize}
The first possibility can be discarded based on geometrical arguments. The quasar is offset from the approximate center of the system, and the clumpy arc deviates in shape and location from a typical partial Einstein ring. However, if this scenario is true, SDSSJ14 would be the most distant gravitational lens ever discovered \citep{2024NatAs...8..119V}, a discovery interesting in its own right. 

Despite the EAZY redshift analysis suggesting a probable spatial correlation between the arc and quasar, and the fact that similar clumpy arc-like formations are typically associated with outbursts of supermassive black holes, the second alternative scenario must be considered. We surveyed the Milliquas quasar catalog \citep{2023OJAp....6E..49F}, searching for all quasars within a one-degree radius around SDSSJ14. Out of 1361 quasars located in this area, around 10\% are expected to be radio-loud quasars with powerful jets \citep{2019A&A...631A..46G}, allowing us to calculate the surface density of jetted quasars. We then derived that the probability for at least one unrelated quasar to be found by chance within 5$\arcsec$ around SDSSJ14 (that is, in the vicinity of the interior side of the arc) is P=2.6$\times$10$^{-4}$.
\section{Summary}
In this paper, we report the discovery of a jet and a hotspot associated with a $z$=2.948 quasar, using archival X-ray and radio data. We uncovered a 250-kpc-long, clumpy, symmetrical arc, oriented perpendicularly to the direction of the jet, using archival optical observations. The analysis of individual clumps suggests that they are likely spatially correlated with each other and with the quasar. Additionally, each clump has masses and star formation rates consistent with those of high-$z$ protogalaxies. We hypothesize that the quasar jet interacts with a nearby cosmic filament, leading to the development of instabilities within the filament and its subsequent fragmentation into galaxy-sized clumps of material - a process never observed at such gargantuan scales before. Depending on their dark matter content, the clumps will either spiral in and feed the central object or form a compact group or the core of a future galaxy cluster.

The properties of SDSSJ14 can only be better constrained with the JWST. Assuming a redshift of $z$=2.948 suggests that the key diagnostic emission lines, such as H$\alpha$, will only be detectable with the JWST due to the atmospheric cutoff at $\geq$2.5 $\mu$m. In addition to confirming the distance and spatial correlation of the arc and the quasar, the emission line analysis could provide information on shocked star formation, accurate masses, and ages of all clumps, transforming our understanding of this system.
\begin{table}[h]
\caption{Properties of selected clumps}\label{tab1}%
\begin{tabular}{@{}llll@{}}
\toprule
ID & $z_{\textup{phot}}$  & log (M$_*$/\(M_\odot\)) & log SFR (\(M_\odot\) yr$^{-1}$)\\

C1    & 2.99$^{+0.13}_{-0.11}$   & 10.73$^{+0.30}_{-0.01}$  & 1.79$^{+0.06}_{-0.11}$  \\
C3    & 2.95$^{+0.26}_{-0.20}$   & 10.29$^{+0.09}_{-0.22}$  & 1.25$^{+0.15}_{-0.09}$  \\
C5    & 3.14$^{+0.13}_{-0.13}$   & 10.19$^{+0.20}_{-0.02}$  & 1.53$^{+0.05}_{-0.10}$  \\
C7 &     3.15$^{+0.18}_{-0.19}$       &  10.24$^{+0.21}_{-0.07}$      &   1.42$^{+0.09}_{-0.10}$         \\
\botrule
\end{tabular}
\end{table}

\bibliography{sample701}{}
\bibliographystyle{aasjournalv7}



\end{document}